\documentclass[%
reprint,
showkeys,
amsmath,amssymb,
aps,
pra,
]{revtex4-2}
\usepackage{graphics}
\usepackage{epsfig}
\usepackage{epstopdf}
\usepackage{amsmath}
\usepackage{array}
\usepackage{subfigure}
\usepackage[colorlinks=true,linkcolor=blue,citecolor=blue,      urlcolor=blue]{hyperref}
\begin{document}
	\title{Modified expansion law with Kodama-Hayward temperature for the horizon}
	\author{Muhsinath. M}
	\email{muhsinathmohideen@gmail.com}
	\author{Hassan Basari V. T.}
	\email{basari@cusat.ac.in}
    \author{Titus K. Mathew}
	\email{titus@cusat.ac.in}
	\affiliation{%
		Department of Physics, Cochin University of Science and Technology, Kochi, Kerala 682022, India
	}%
	\begin{center}
		\begin{abstract}
The expansion law proposed by Padmanabhan 
suggests that the evolution of the volume of the horizon is due to the difference between the degrees of freedom on the horizon and the degrees of freedom in the bulk enclosed by the horizon. In formulating this law,  Padmanabhan used the temperature, $ T=H/2\pi $ for a dynamical expansion.
In this work, we modified the expansion law using Kodama-Hayward temperature, the dynamical temperature, for the horizon, first in (3+1) Einstein's gravity and extended it to high order gravity theories such as (n+1) Einstein gravity, Gauss-Bonnet gravity, and more general Lovelock gravity. Contrary to the conventional approach, we expressed degrees of freedom of the horizon in terms of the ‘surface energy' of the horizon. Also, we have expressed modified expansion law in terms of cosmic components. It then turns out that, it is possible to express the modified expansion law in a form as if $T=H/2\pi$ is the temperature of the dynamical horizon. 
		\end{abstract}

		\maketitle
		
\end{center}

\section{Introduction}
Recent research put forward an intriguing possibility that gravity could be an emergent phenomenon like elasticity or fluid mechanics. In the past, Sakharov \cite{1}, in 1968, gives the first hint in this direction, in which gravity is considered to be arisen due to the mean field approximation of some underlying degrees of freedom. The idea of emergent gravity was resurrected recently, following the establishment of the intriguing correlation between gravity and thermodynamics. Following the laws of black hole mechanics, by Bekenstein, Bardeen, and Hawking \cite{2,3,4}, it was established that a black hole horizon can be associated with an entropy proportional to its area and a temperature proportional to its surface gravity. This idea was later extended to the cosmological horizon of de Sitter universe by Gibbons and Hawking \cite{5}. Meantime, in 1995, Jacobson  \cite{6} derived Einstein field equation from the Clausius relation $ \delta Q= TdS $, where  $ \delta Q $ is the energy flux across the horizon, $ T $ is the Unruh temperature seen by an accelerating observer inside the horizon and $S$ is the Bekenstein entropy of the horizon. In 2010, Padmanabhan \cite{7,8} obtained Newton's law of gravity by combining the equipartition law of gravitating energy within the volume bounded by the horizon and the thermodynamic relation
$ S = (1/2)E/T $, where $ S $ is the horizon entropy, $ E $ is the effective gravitational mass enclosed by the horizon, and $ T $ is
the temperature of the horizon. On the other hand, Verlinde argued that gravity could be an entropic force caused by changes in the information associated with the positions of material bodies, and using this idea the author derived Newton's law of gravitation \cite{9}. However, later some authors \cite{10,11,12,13,14,15} have opened that, the basic idea of Verlinde may not be correct.

Following the emergent gravity perspective,  Padmanabhan introduced that cosmic space can also be an emergent entity. According to this, the expansion of the universe can be thought of as the emergence of cosmic space as cosmic time progress. Conceptually it is difficult to consider time as emerging from some pre-geometric variables. 
But this inconsistency will disappear in cosmology, where all geodesic observers measure the same time, the proper time, and for whom CMBR appears homogeneous and isotropic \cite{16}. With the emergence paradigm, Padmanabhan proposed a fundamental law, known as the law of emergence, which describes the evolution of the volume of the horizon happened due to the difference between the degrees of freedom on the surface of the horizon and that in bulk enclosed by the horizon. 
Padmanabhan \cite{16} derived the Friedmann equation of flat FRW universe in (3+1) Einstein gravity from the law of emergence. 

Padmanabhan's proposal was extended by Cai \cite{17} to (n+1) Einstein gravity, Gauss-Bonnet gravity, and more general Lovelock gravity, for flat FRW universe, by modifying the degrees of freedom on the surface of the horizon. An extension of the law to the non-flat universe was done by  Sheykhi \cite{18} by appropriately modifying the expressions for degrees of freedom. An alternative approach for the expansion law was given by Yang et al.\cite{19}, where the authors have assumed the increase in the horizon volume as a function of the difference in the degrees of freedom for a flat FRW universe. They consider the difference in the degrees of freedom as a Taylor series of the function $f(\Delta N, N_{sur}),$ which is truncated at the first order. 

In reference \cite{20}, authors derived the expansion law from the first law of thermodynamics and extended it to higher order gravity theories such as (n+1) Einstein gravity, Gauss-Bonnet gravity, and more general Lovelock gravity, for both flat and non-flat FRW universe by suitably modifying surface degrees of freedom in the respective theories. The equivalency between holographic equipartition and maximization of entropy has been proved in reference \cite{21}. In 2019, Krishna and Mathew \cite{22} have shown that the law of emergence implies the maximization of horizon entropy in the context of Einstein's gravity. This result has been extended to the higher order gravity theories like Gauss-Bonnet, and more general Lovelock theories of gravity by the same group \cite{21,22,23}.  

In most of the previous attempts, authors have used the temperature of the dynamical horizon as $ T=H/2\pi$ to obtain the degrees of freedom within the bulk enclosed by the horizon. This relation for the temperature of the cosmological horizon is an extension of the temperature relation for the black hole horizon \cite{24}. In fact, the temperature of the horizon is proportional to its surface gravity. Accordingly, the temperature of the dynamical horizon of a flat universe can be expressed as $T=-H/2\pi(1+\dot H/2H^2)$ \cite{25}. Despite this, the use of the temperature of the form $T=H/2\pi$ for the horizon has been justified by the respective authors mainly by pointing out the simplicity of the resulting expansion law, which anyway implies the correct Friedmann equations. 
For instance, in the original proposal by Padmanabhan, it was argued that the temperature, $T=H/2\pi$ is a direct extension of the black hole horizon temperature to the cosmological horizon, hence can be used for the dynamical Hubble horizon too, and it leads to a simple form for the law of expansion, which nicely reduces to the Friedmann equations.
He also mentioned that defining horizon temperature in a non-de Sitter universe when $ H $ is time-dependent is not easy since there exists no proper Killing vector to define the surface gravity. So some controversy in the literature regarding the correct choice of temperature persists \cite{16}. However, it pointed out that
in its strict sense, the temperature of the form $T=H/2\pi$ is true only for the de Sitter horizon for which the Hubble parameter becomes a constant. Hence, taking the temperature of a dynamical horizon the same as that of a static horizon is unreasonable.

It is to be noted that the temperature can be precisely defined for a horizon only if there exists a proper Killing vector, which does exist for a de Sitter horizon. However, there exists no timelike Killing vector for a cosmological dynamic spacetime which makes the definition of the temperature of such a horizon cumbersome. A solution to this was proposed by Kodama \cite{26}, by which one could define what is known as the Kodama vector, which has almost all the properties of the Killing vector. The Kodama vector generates a preferred flow of time and it is a dynamic analogue of a stationary Killing vector that exists for de Sitter horizon \cite{27}. It was shown by Hayward \cite{28}, that a temperature based on the Kodama vector can be defined for the horizon of a dynamical black hole.
In the case of cosmology, 
this temperature, which can be called as Kodama-Hayward temperature, contained, apart from $H/2\pi,$ terms proportional to the derivative of Hubble parameter in the case of a flat universe \cite{26,27,28,29,30,31,32,33,34,35}.  

Unlike the previous approaches, in reference \cite{27}, Dezaki and Mirza have derived modified expansion law for a Kodama observer from the unified first law of thermodynamics using the Kodama-Hayward temperature. But their work has the following issues, (1) the law will not reduce to the original law proposed by Padmanabhan for the flat universe in the appropriate limit \cite{27}, and (2) It is not possible to extend it to more general gravity theories, like Gauss-Bonnet, and Lovelock \cite{20}. In the present work, we first derive a proper form for the expansion law from the first law of thermodynamics, using the Kodama-Hayward temperature. We then extend the result to (n+1) Einstein gravity, Gauss-Bonnet gravity, and more general Lovelock gravity. Another point to be noted, unlike the previous approaches \cite{17,18,20,27}, we define the surface degrees of freedom (on the horizon), in terms of temperature, like the bulk degrees of freedom. The advantage of such a definition of the surface degrees of freedom is that it brings a  unified method to define both degrees of freedom for any gravity theory. While in the previous works the authors were forced to modify the definitions of the surface degrees of freedom on the horizon subjected to the gravity theory under consideration \cite{17,18,20}. 

In this paper, we will derive modified expansion law from the unified first law of thermodynamics using Kodama-Hayward temperature for the horizon, which is the correct temperature for a dynamic apparent horizon \cite{26,27,32}. In section II, we will define a new relation for surface degrees of freedom at the horizon and also derive the standard Friedmann equation from the modified expansion law. In section III, we will derive the modified expansion law in (n+1) Einstein gravity. We extended our study to higher order gravity theories in section IV. In section V, we expressed modified expansion law in terms of the equation of state of the cosmic components to study our results further. 
\section{Expansion law from first law of thermodynamics using Kodama-Hayward temperature}
Following the work of Hayward \cite{28,32}, the temperature of the dynamic apparent horizon for the Kodama observer is related to the surface gravity of the horizon. Accordingly, the temperature, called Kodama-Hayward temperature, can be written as,
\begin{equation}
	\label{1}
	T=\dfrac{\kappa}{2\pi}= -{\frac{1}{2\pi\tilde{r}_{A}}}\left(1-\epsilon\right)
\end{equation}
where $\kappa$ is the surface gravity of the horizon and $\tilde{r}_A$ is the radius of the apparent horizon. 
Here, 
\begin{equation}
	\label{2}
	\epsilon=\frac{\dot{\tilde{r}}_{A}}{2H\tilde{r}_{A}}
\end{equation}
where over-dot represents the derivative with cosmic time.
For a flat universe, the apparent horizon becomes the Hubble horizon with $\tilde{r}_{A} \sim \frac{1}{H} $, for which the temperature \eqref{1} takes the form,
\begin{equation}
	\label{3}
	T=-{\frac{H}{2\pi}}\left(1+\frac{\dot{H}}{2H^2}\right)
\end{equation}
We now aim to derive the expansion law from the unified first law of thermodynamics using Kodama-Hayward temperature for the apparent horizon. 
The unified first law is of the form \cite{29,32,33},
\begin{equation}
	\label{4}
	dE=TdS+WdV
\end{equation}
where $E$ is the total energy of the matter inside the horizon,
$	W=(\rho-P)/2, $ is the work density, with $\rho$ and $P$ as the density and pressure of the cosmic fluid, 
$	V= \frac{4}{3} \pi {\tilde{r}_{A}}^{~3} $ is the volume of the apparent horizon 
and $ S $ is the entropy of the horizon given by,
\begin{equation}
	\label{5}
	S=\frac{A}{4L_{P}^2}
\end{equation}
which is the Bekenstein formula for the entropy of the horizon. Using the temperature of the horizon \eqref{1} and entropy of the horizon \eqref{5}, we can express the 
the first term on the right hand side of the unified first law as,
\begin{equation}
	\label{6}
	TdS=-(1-\epsilon)AH{\tilde{r}_{A}}(\rho+P)dt
\end{equation}
Similarly, using work density relation and volume of the apparent horizon, we get the second term on the right hand side of the unified first law as,
\begin{equation}
	\label{7}
	WdV=\epsilon AH\tilde{r}_{A}(\rho-P)dt
\end{equation}
Then by the unified first law, the change in energy within the horizon, which is equal to the energy crossing the apparent horizon within the interval $ dt $ can be obtained by adding \eqref{6} and \eqref{7}, which on suitable rearrangement leads to \cite{27,29},
\begin{equation}
	\label{8}
	\frac{dE}{dt}= -H \left( \left(\rho+3P\right)V+2\rho V-6\epsilon \rho V\right)
\end{equation}
From the above equation, the Komar energy can be written as \cite{18,19,27},
\begin{equation}
	\label{9}
	(\rho+3P)V=-\frac{1}{H}\frac{dE}{dt}-2\rho V(1-3\epsilon)
\end{equation}
Here $E=\rho V,$ from which it can be shown that, 
\begin{equation}
		\label{10}
		\frac{dE}{dt}=-AH\tilde{r}_{A}((1-2\epsilon)\rho+P)
	\end{equation}
which on substituting into equation \eqref{9} and simplifying it using the result \eqref{6}, we arrive at,
\begin{equation}
	\label{11}
	(\rho +3P)V=-\frac{T}{H(1-\epsilon)}\frac{dS}{dt}-2\rho V
\end{equation}
Substituting the temperature from equation \eqref{1}, and rearranging the above equation, we get,
\begin{equation}
	\label{12}
	\frac{1}{2\pi \tilde{r}_{A}H}\frac{dS}{dt}=(\rho +3P)V+2\rho V
\end{equation}
The first term on the right hand side of the above equation is the Komar energy enclosed by the horizon. In line with this, let us define the second term on the right hand side as the energy associated with the surface of the horizon, i.e.,
\begin{equation}
	\label{13}
	E_{sur}= 2\rho V
\end{equation}
Using this equation \eqref{12} can suitably transformed to relation for $dV/dt.$ For this substitute for entropy \eqref{5} and the standard result $dA/dV=2/\tilde{r}_A$ into the relation \eqref{12}. The resulting equation will be multiplied throughout with $2/T,$ and finally arrive at, 
\begin{equation}
	\label{14}
	\frac{dV}{dt}=\frac{1}{2} TAHL_{P}^2(N_{sur}-N_{bulk})
\end{equation}
which is the law of expansion.
Here we have identified the bulk  degrees of freedom enclosed by the horizon and degrees of freedom on the surface of the horizon as,
\begin{equation}
	\label{15}
	N_{bulk} = -\frac{2(\rho+3P)V}{T}, \quad N_{sur}=\frac{2E_{sur }}{T}
\end{equation}
Here the negative sign in the expression for $N_{bulk}$ guarantees the positiveness of the bulk degrees of freedom during the accelerating epoch, at which $(\rho+3P)<0$ and $ E_{sur}$ is given by \eqref{13}. Here the novel fact is that the surface degrees of freedom has been identified as the ratio of the surface energy and the temperature. This form for $N_{sur}$ is conceptually different from its original definition, where it is defined as the number of Planck area on the horizon surface. However, on substituting $\rho, V$ in the above relation, and taking the temperature as $T=H/2\pi,$ the above expression for the surface degrees of freedom will be reduced to, 
$N_{sur}=A/L_P^2,$ equivalent to the enumeration of the Planck area on the horizon surface which is exactly the same as the original definition used by Padmanabhan. To be more specific, the advantage of the above definition of $N_{sur}$ is that, it defines the surface degrees of freedom 
in terms of temperature, and hence is symmetrical with the definition of the bulk degrees of freedom. 

Let us now contrast equation \eqref{14} with the original equation for the law of emergence proposed by Padmanabhan. In the original proposal, the rate of change of the Hubble volume is proposed to be equal to $L_P^2 (N_{sur}-N_{bulk}).$  However, in the present case, when we used the Kodama-Hayward temperature, it turns out that, the rate of change of horizon volume is proportional to:
\begin{enumerate}
	\item the difference in degrees of freedom, $(N_{sur} - N_{bulk})$
	\item  the temperature of the horizon, $T$ 
	\item the Hubble parameter, $H$
	\item the area of the horizon, A
\end{enumerate}
Soon we will show that the modified law of expansion will imply the corresponding Friedmann equations. Apart from that, an interesting consequence is the implication of the holographic equipartition condition 
$ N_{sur}=N_{bulk}, $ which is satisfied asymptotically by the universe. This condition can be expressed as,
\begin{equation}
	\label{16}
	\frac{2\rho V}{(1/2)T}= \frac{|(\rho+3P)|V}{(1/2)T}
\end{equation}
when $\rho=-P$ which corresponds to the end de Sitter epoch.
This immediately implies 
the equality of the surface energy and volume energy, which is the Komar energy, as
\begin{equation}
	\label{17}
	E_{sur}= |E_{Komar}|
\end{equation}
This shows that, the equality of the degrees of freedom directly at the end stage implies the equality of the surface and effective volume energy of the horizon of the expanding universe. It should be noted that when one adopts an approximate relation of temperature, as  $ T \approx \frac{H}{2\pi}, $ and subject to the condition that, $\epsilon << 1,$ then the modified law of emergence given in equation \eqref{14}, reduces to the original law due to Padmanabhan. 

Now we will show that the modified law of expansion will lead to the standard Friedmann equation. On substituting the degrees of freedoms in equation \eqref{15}, and also using equation \eqref{13}, into the modified expansion law \eqref{14} we get,
\begin{equation}
	\label{18}
	\frac{dV}{dt}= AHL_{P}^2\left. [2\rho+(\rho+3P)]\right. V
\end{equation}
The rate of change of the apparent horizon volume in the left hand of the above equation is proportional to the time derivative of its radius, 
\begin{equation}
	\label{19}
	\dot{\tilde{r}}_{A}=-H\tilde{r}_{A}{^3}\left(\dot{H}-\frac{k}{a^2}\right)
\end{equation}
The modified law of  emergence then takes the form, 
\begin{equation}
	\label{20}
	-\left(\dot{H}-\frac{k}{a^2}\right)=\frac{4\pi L_{P}^2}{3}\left.[3(\rho+P)]\right.
\end{equation}
Multiplying the above equation by $ 2H $ and then 
using the continuity equation $\dot{\rho}+3H(\rho+P)=0$, we can have,
\begin{equation}
	\label{21}
	\frac{d}{dt} \left(H^2+\frac{k}{a^2}\right)= \frac{8\pi L_{P}^2}{3}\left. \frac{d\rho}{dt}\right. 
\end{equation}
The above equation upon integration gives,
\begin{equation}
	\label{22}
	H^2+\frac{k}{a^2}=\frac{8\pi L_{P}^2}{3}\rho      
\end{equation}
This equation is nothing but the Friedmann equation in a non-flat FRW universe. 

At this juncture, let us compare our results with that
in reference \cite{27}, in which the authors derived the expansion law for a Kodama observer in (3+1) Einstein gravity, by assuming the Misner–Sharp energy in the first law of thermodynamics, as $E_{MS}=2TS+3WV.$ 
They have used the surface degrees of freedom as $ A/L_{P}^2, $
and the bulk degrees of freedom as $ N_{bulk}=\frac{2(\rho+3P)V}{T},$ dropping the negative sign as in the conventional definition. The disadvantage of their formalism is that the derived law of expansion will not reduce to the original law by Padmanabhan, in the limiting condition, $\epsilon <<1,$ with temperature of the horizon as $T=H/2\pi.$ Moreover, it is impossible to extend their formalism to higher gravity theories due to the non-proper choice for the Misner-Sharp energy \cite{29}.

\section{Modified expansion law in (n+1) Einstein gravity}
In this section, we will extend the above procedure 
to generalize the law of expansion to (n+1) dimensional FRW universe in the context of Einstein's gravity. 
The entropy of the horizon in (n+1) dimensional universe given by, 
\cite{18,20}
\begin{equation}
	\label{23}
	S=\frac{\mathcal{A}}{4L_{P}^{n-1}} 
\end{equation}
where $ \mathcal{A}=n\Omega_{n} \tilde{r}_{A}^{n-1} $ is the area of the apparent horizon with volume $ V= \Omega_{n} \tilde{r}_{A}^{n} $ of (n+1) dimensional FRW universe. Now consider the unified law in equation \eqref{4}. As in the previous section substitute the temperature of the horizon \eqref{1}, the entropy of the horizon \eqref{23}, 
and 
work density, and volume of the apparent horizon 
we get, after some re-arrangements, 
\begin{equation}
	\label{24}
	\frac{1}{2\pi \tilde{r}_{A}H}\frac{dS}{dt}=[(n-2)\rho+nP]V+2\rho V
\end{equation}
Substituting for entropy from equation \eqref{23} and then using $ d\mathcal{A}/dV=(n-1)/\tilde{r}_{A} $ we get,
\begin{equation}
	\label{25}
	\frac{(n-1)}{2AHL_{P}^{n-1}}\frac{dV}{dt}=[(n-2)\rho+nP]V+2\rho V
\end{equation}
where $ A $ is the area of the apparent horizon in three dimensional universe.
Now,
multiplying on both sides of the above equation by $ \alpha=(n-1)/2(n-2) $ gives,
\begin{equation}
	\label{26}
	\frac{\alpha}{2AHL_{P}^{n-1}}\frac{dV}{dt}=\frac{1}{2}\left[\frac{[(n-2)\rho+nP]V}{(n-2)}+\frac{2\rho V}{(n-2)} \right] 
\end{equation}
Let us define the surface energy of the horizon in the (n+1) dimensional FRW universe as,
\begin{equation}
	\label{27}
	E_{sur}= \frac{2\rho V}{(n-2)}
\end{equation}
Substituting \eqref{27} and then multiplying on both sides of \eqref{26} by $ 2/T$ we get,
\begin{equation}
	\label{28}
	\alpha	\frac{dV}{dt}=\frac{1}{2}TAHL_{P}^{n-1}(N_{sur}-N_{bulk})
\end{equation}
The above equation is the modified expansion law in (n+1) Einstein gravity with Kodama-Hayward temperature for the horizon.
Here, we identify surface and bulk degrees freedom of horizon in (n+1) dimensional FRW universe as, 
\begin{equation}
	\label{29}
	N_{bulk}=-\frac{[2(n-2)\rho+nP]V}{(n-2)T}, \quad  N_{sur}=\frac{2E_{sur}}{T}
\end{equation}
respectively.
Now, let's consider another approach where we do not multiply equation \eqref{25} by $ \alpha $ and use the surface energy of the horizon 
same as (3+1) dimensional universe.
The resulting equation, in that case, is then multiplied throughout with $ 2/T $, which gives modified expansion law as,
\begin{equation}
	\label{30}
	\frac{dV}{dt}=\frac{1}{(n-1)}TAHL_{P}^{n-1}(N_{sur}-N_{bulk})
\end{equation}
by identifying degrees of freedom in (n+1) dimensional universe as,
\begin{equation}
	\label{31}
	\begin{split}
		N_{bulk}&=-\frac{[2(n-2)\rho+nP]V}{T}, \\  N_{sur}&=\frac{2E_{sur}}{T}, \quad E_{sur}=2\rho V
	\end{split}
\end{equation}
There is no $ \alpha $ term in the above modified expansion law \eqref{30}.
Both the laws \eqref{28} and \eqref{30} reduce to the law due to Padmanabhan when $ T=H/2\pi$ for a (3+1) dimensional FRW universe. 
But in the first equation the surface degrees of freedom \eqref{29} takes form $N_{sur}=\alpha A/L_{P}^{n-1}$ 
while from the second equation \eqref{31}, the surface degrees of freedom turns out to be just $ N_{sur}=A/L_{P}^{n-1}$ when $ T=H/2\pi $.

Compared to the corresponding equations of the law of expansion obtained by Cai and Sheykhi \cite{17,18}, the above equations, \eqref{28} or/and \eqref{30}, the emergence of space depends on, apart from $(N_{sur}-N_{bulk}),$ temperature, area of the horizon, Hubble parameter. Moreover, the surface degrees of freedom defined as the ratio of the surface energy and the temperature.

\section{Modified expansion law in Gauss-Bonnet and Lovelock gravity}
In this section, we will derive the law of expansion using the Kodama-Hayward temperature of the horizon for more general gravity theories. 
It is known that the entropy formula of the black hole in the Gauss-Bonnet gravity is not just proportional to the area of the horizon instead it contains additional correction terms, such that \cite{18,20}, 
\begin{equation}
	\label{32}
	S=\frac{A_{+}}{4L_{P}^{n-1}}\left[1+\frac{n-1}{n-3} \frac{2\tilde{\alpha}}{{r}_{+}^{2}} \right] 
\end{equation}
where $ A_{+}=n\Omega_{n} \tilde{r}_{+}^{n-1} $ is the area of the black hole horizon and $ r_{+} $ is the the radius of the black hole horizon. Also, $\tilde{\alpha}=(n-2)(n-3)\alpha$ is the Gauss-Bonnet coefficient \cite{18,20}. Assuming that the entropy formula in equation \eqref{32} also holds for the apparent horizon in FRW universe, we get the form of entropy for the apparent horizon in Gauss-Bonnet gravity by replacing black hole horizon radius $ r_{+} $ by apparent horizon radius $\tilde{r}_{A}$ as \cite{17,18,20,29,30},
\begin{equation}
	\label{33}
	S=\frac{\mathcal{A}}{4L_{P}^{n-1}}\left[1+\frac{n-1}{n-3} \frac{2\tilde{\alpha}}{\tilde{r}_{A}^{2}} \right] 
\end{equation}
where $ \mathcal{A}=n\Omega_{n} \tilde{r}_{A}^{n-1} $ is the area of the apparent horizon \cite{18,20}. The effective area corresponding to entropy \eqref{33} is given by,
\begin{equation}
	\label{34}
	\tilde{A}=\mathcal{A}\left[1+\frac{n-1}{n-3} \frac{2\tilde{\alpha}}{\tilde{r}_{A}^{2}} \right] 
\end{equation}
The increase in effective area is given by \cite{17,18},
\begin{equation}
	\label{35}
	\frac{d\tilde{A}}{dt}=\frac{(n-1)}{\tilde{r}_{A}} \frac{d\tilde{V}}{dt}
\end{equation}
Following the same steps as in the previous section, one can then express the unified first law of thermodynamics, as an equation for 
$ d\tilde{V}/dt. $ 
The resulting equation is then multiplied on both sides by $ \alpha $ and $ 2/T $, which gives modified expansion law in Gauss-Bonnet gravity as,
\begin{equation}
	\label{36}
	\alpha	\frac{d\tilde{V}}{dt}=\frac{1}{2}TAHL_{P}^{n-1}(N_{sur}-N_{bulk})
\end{equation}

Now we will obtain the law of expansion in the context of Lovelock gravity. Consider the entropy formula for the spherically symmetric black hole in more general Lovelock gravity,
\begin{equation}
	\label{37}
	S=\frac{A_{+}}{4L_{P}^{n-1}}\sum_{i=1}^{m}\frac{i(n-1)}{(n-2i+1)} \hat{c}_{i}{r}_{+}^{2-2i}
\end{equation}
where $ m=[n/2] $ and the coefficients $\hat{c}_{i} $ are given by,
\begin{equation} 
	\label{38}
	\begin{split}
		\hat{c}_{0}& =\frac{c_{0}}{n(n-1)}, \quad \hat{c}_{1}=1\\
		\hat{c}_{i}&=c_{i}\prod_{j=3}^{2m} (n+1-j), \quad i>1
	\end{split}
\end{equation}
This relation for entropy can now be adopted for the apparent horizon $\tilde{r}_{A}, $ and hence 
the entropy of horizon in Lovelock gravity is \cite{18},
\begin{equation}
	\label{39}
	S=\frac{\mathcal{A}}{4L_{P}^{n-1}}\sum_{i=1}^{m}\frac{i(n-1)}{(n-2i+1)} \hat{c}_{i}\tilde{r}_{A}^{2-2i}
\end{equation}
The effective area of the apparent horizon in Lovelock gravity is given by \cite{18,20},
\begin{equation}
	\label{40}
	\tilde{A}=\mathcal{A}\sum_{i=1}^{m}\frac{i(n-1)}{(n-2i+1)} \hat{c}_{i}\tilde{r}_{A}^{2-2i}
\end{equation}
Increase in the effective area in the present case follows a similar relation as given by \eqref{35}.
Following the same steps as in Gauss-Bonnet gravity and using entropy \eqref{39} we finally get modified expansion law in Lovelock gravity as exactly similar in form as that in \eqref{36}. 

It is to be noted that equations for surface energy of the horizon and degrees of freedom of the horizon in Gauss-Bonnet gravity and Lovelock are the same as those expressions in (n+1) dimensional FRW universe.
It is possible to obtain, the law of expansion, without 
multiplying the left side of the equation \eqref{25} by $ \alpha $ after substituting for entropy.
Then we get modified expansion law in Gauss-Bonnet gravity and Lovelock gravity as,
\begin{equation}
	\label{41}
	\frac{d\tilde{V}}{dt}=\frac{1}{(n-1)}TAHL_{P}^{n-1}(N_{sur}-N_{bulk})
\end{equation}
with degrees of freedom of the horizon same as in \eqref{31}.
The important fact compared to the previous works \cite{17,18,20,27} in this regard is that, here, we extended the modified expansion law in (n+1) Einstein gravity to Gauss-Bonnet and more general Lovelock gravity without modifying equations of $ N_{sur} $ and $ N_{bulk} $ in those theories.
\section{Modified expansion law 
	and the cosmic components}
One of the issues, that seems to arise at first look into
the law of expansion in equation \eqref{14}, 
is the occurrence of singularity in degrees of freedom when $\epsilon = 1,$ at which the temperature becomes zero. However, the temperature, $T$ that appears as a coefficient on the right hand side of the law of expansion \eqref{14} will effectively cancel the same temperature that appears in the denominator of the degrees of freedoms, $N_{sur}$ and $N_{bulk}.$ Hence this apparent singularity has been effectively nullified in the law of expansion. This is the prime advantage of our definition of both degrees of freedom in the equipartition way.  

Now consider the law of expansion \eqref{14}. As we mentioned before, the temperature in the coefficient will get cancelled with that in the denominator of the expressions for $N_{sur}$ and $N_{bulk}.$ The resulting expression will be,
\begin{equation} 
	\label{42}
	\frac{dV}{dt}=AHL_P^2 \left(E_{sur} + E_{bulk} \right)
\end{equation}
where we have substituted for the degrees of freedom from equations \eqref{15}. Into this equation, we will first substitute both $E_{sur}$ from equation \eqref{13} and $E_{bulk}$ as the Komar energy.  
Let us now restrict ourselves to a flat universe for convenience.
We will now replace the pressure in the resulting equation using the equation of state $ P=\omega\rho.$ The law of expansion will now reduce to a simple form,
%
\begin{equation}
\label{43}
\frac{dV}{dt}=\frac{3}{2}(1+\omega)A
\end{equation}
The equation of state parameter, $\omega$ can take values zero for normal matter (and cold dark matter), $1/3$ for radiation, and $-1$ for the cosmological constant. 
For radiation, matter, and dark energy dominated universe rate of change of Hubble volume of the horizon becomes 	
$ 2A ,$
$ 3A/2, $
and	$ 0 $
respectively by applying the corresponding equation of state parameter $ \omega $ according to \eqref{45}.
This, in turn, implies that the rate of change in Hubble volume decreases from the early radiation epoch onward and becomes zero at the late dark energy dominated epoch. At the end de Sitter epoch, the Hubble volume becomes a constant.
	
Now consider equation \eqref{42}. In the product, $AH=4\pi/H.$ On taking this inside the parenthesis, equation \eqref{42} can be re-expressed as,
\begin{equation}
\label{44}
\frac{dV}{dt}=L_P^2 \left(\frac{2E_{sur}}{H/2\pi} + \frac{2E_{bulk}}{H/2\pi} \right)
\end{equation} 
for a flat universe in the context of Einstein's gravity. Now the first term inside the parenthesis of the above equation can be taken as the degrees of freedom on the horizon surface, $N_{sur}$ with temperature $H/2\pi$ and the second term can be taken as the negative the bulk degrees of freedom, $N_{bulk}$ in equipartition with the temperature $H/2\pi.$ Consequently the above equation will be reduced to the form,
\begin{equation}
		\label{45}
		\frac{dV}{dt}=L_P^2 \left(N_{sur} - N_{bulk} \right)
\end{equation}
which is nothing but the exact original form of expansion law proposed by Padmanabhan in the context of the flat universe in Einstein's gravity, where the temperature of the horizon is taken as $T=H/2\pi.$ This is of particular interest since we have started with the Kodama-Hayward temperature. But for flat universe, it turns out that, the law of expansion naturally adopts the temperature $H/2\pi$ for the horizon, such that the equation reduces to the original form.
The same results will follow in the case of non-flat universe also for which 
the resulting expansion law will take the form,
the use of the temperature $ T=1/2\pi\tilde{r}_{A}, $ for the dynamical horizon can be justified, while one forming the law of expansion \cite{18}. It is to be noted that, the temperature relation doesn't depend on the gravity theory, hence a justification of using this temperature for defining the expansion law in higher order gravity theories like Gauss-Bonnet, Lovelock, etc. can similarly be obtained. At this juncture, it is to be noted that, 
there exists some amount of controversy in the literature regarding the correct choice of temperature \cite{16}. Padmanabhan, while using the temperature of the horizon as $ H/2\pi, $ for defining the bulk degrees of freedom, argued that by using this temperature the resulting expansion law has a simple form, and also one can nicely derive the Friedmann equation for a flat universe. In the present work, we started with the Kodama-Hayward temperature for defining both degrees of freedom to formulate the expansion law. It turns out that, the law will suitably reduce to a form, where the temperature of the horizon is, $T=H/2\pi,$ for flat universe.
\section{Conclusions}
Following the emergent gravity paradigm,  Padmanabhan proposed the law of expansion to explain the expansion of the universe, which views that, the evolution of the volume of the horizon is driven by  
the difference in the degrees of freedom on the horizon and that in the bulk enclosed by the horizon. Originally the surface degrees of freedom was defined as the number of Plank area on the surface of the horizon and 
bulk degrees of freedom, as the ratio of the gravitational energy to the temperature $ T=H/2\pi $ of the horizon. This idea was further investigated by many others by using the same temperature \cite{17,18,19,20}. Many have argued that the temperature used is actually the temperature of the horizon of the universe in the de Sitter epoch. Hence it necessitates using the temperature which takes account of the dynamical evolution of the temperature as the universe expands.

In this work, we derived a modified expansion law from the unified first law of thermodynamics using Kodama-Hayward temperature \eqref{1} for the horizon, which accounts for the dynamical evolution of the horizon. The law thus obtained in equation \eqref{14}, has a marked difference compared to the original form. The rate of change of the volume of the apparent horizon, is proportional, apart from the difference between degrees of freedom, to the Hubble parameter, the temperature of the horizon, and area of the horizon. It reduces to the original law due to Padmanabhan in the appropriate limit. One of the novel thing in present form of the law is that, we could define the surface degrees of freedom as the ratio of the surface energy of the horizon to its temperature, which is contrary to the conventional approach. 

We have extended our results to (n+1) Einstein's gravity and also to higher order gravity theories, like Gauss-Bonnet and Lovelock. It is to mention that in all these extensions we 
obtained the degrees of freedom, both surface and bulk, as the ratio of the respective energy and the horizon temperature.
As a further check, we have shown it is possible to obtain the Friedmann equations of the expanding universe of any spatial curvature from the modified expansion. 

We expressed the modified law in terms of the equation of the state of cosmic components, after substituting for modified degrees of freedom of the horizon. Consequently, it appears that the modified law naturally appears to favor the temperature, $T= H/2\pi$ for the horizon. Thus finally it turns out that, the present approach of modifying the law of expansion by using dynamical temperature for the horizon, upholds the original proposal of 
Padmanabhan that, the choice of the temperature $T=H/2\pi,$ is still admissible with the proper form of the law. 
\section*{Acknowledgments}
Authors thank P. B. Krishna and M. Deepika for discussions. Muhsinath. M acknowledges KSCSTE, Government of Kerala for Prathibha Scholarship.

\bibliographystyle{apsrev4-1}
\bibliography{ref}
\end{document}